\documentclass{egpubl}
\usepackage{eurovis2023}
\ConferencePaper        
\usepackage[T1]{fontenc}
\usepackage{dfadobe}  
\usepackage{cite}  
\BibtexOrBiblatex
\electronicVersion
\PrintedOrElectronic
\ifpdf \usepackage[pdftex]{graphicx} \pdfcompresslevel=9
\else \usepackage[dvips]{graphicx} \fi
\usepackage{egweblnk}

\usepackage{soul}
\usepackage{booktabs}
\usepackage{graphicx}
\usepackage{float}

\newcommand{\red}[1]{\textcolor{black}{#1}}

\title[LINGO : Visually Debiasing Natural Language Instructions to Support Task Diversity]%
      {LINGO : Visually Debiasing Natural Language Instructions\\to Support Task Diversity}

\author[Arunkumar et al.]
{\parbox{\textwidth}{
\centering A.\,Arunkumar$^{1}$,
        S. Sharma$^{1}$,
        R. Agrawal$^{1}$,
        S. Chandrasekaran$^{1}$ 
        and C. Bryan$^{1}$
        }
        \\
    {
    \parbox{\textwidth}{\centering $^1$Arizona State University, Tempe, United States}
    }
}

\begin{document}
\teaser{
 \includegraphics[width=0.7\linewidth]{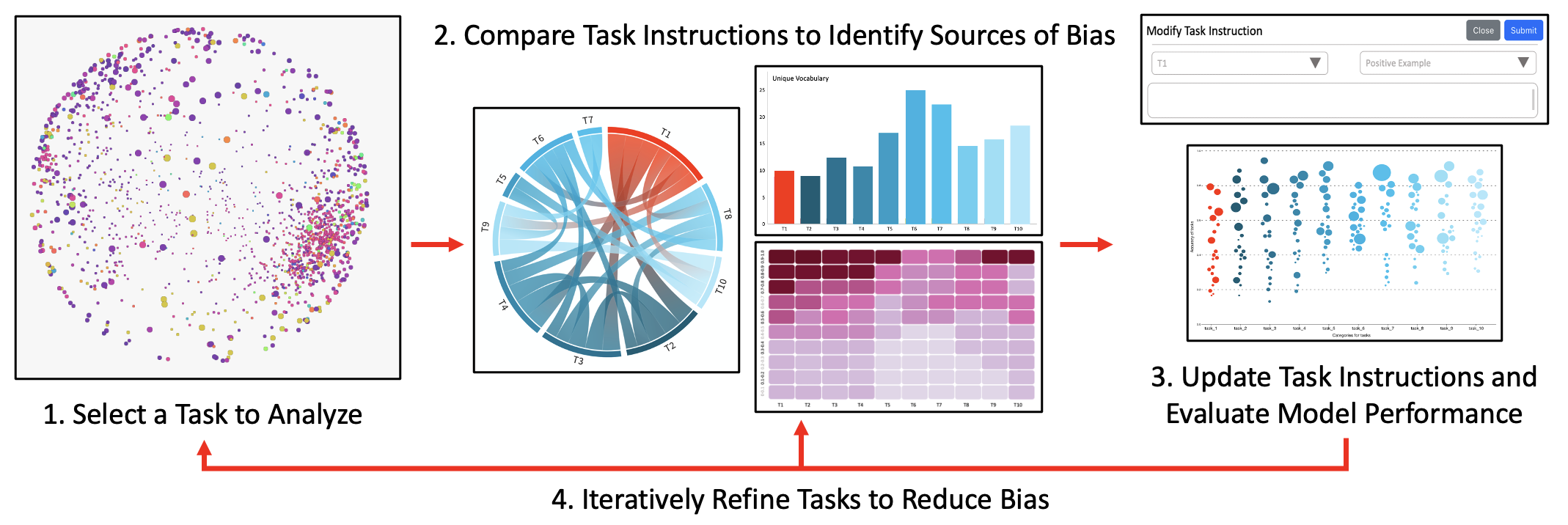}
 \centering
  \caption{LINGO workflow: (1) Select a task to analyze based on the sentence embedding space of task instructions. (2) Next, compare the  linguistic characteristics of nearest neighbor task instructions to identify features that lead to instruction bias. (3) Modify or create task instructions, and evaluate model performance on the updated task. (4) Tasks can be iteratively modified till bias is sufficiently reduced.}
\label{fig:teaser}
}
\maketitle
\begin{abstract}
Cross-task generalization is a significant outcome that defines mastery in natural language understanding. Humans show a remarkable aptitude for this, and can solve many different types of tasks, given definitions in the form of textual instructions and a small set of examples. Recent work with pre-trained language models mimics this learning style: users can define and exemplify a task for the model to attempt as a series of natural language prompts or instructions. While prompting approaches have led to higher cross-task generalization compared to traditional supervised learning, analyzing `bias' in the task instructions given to the model is a difficult problem, and has thus been relatively unexplored. For instance, are we truly modeling a task, or are we modeling a user's instructions? To help investigate this, we develop LINGO, a novel visual analytics interface that supports an effective, task-driven workflow to (1) help identify bias in natural language task instructions, (2) alter (or create) task instructions to reduce bias, and (3) evaluate pre-trained model performance on debiased task instructions. To robustly evaluate LINGO, we conduct a user study with both novice and expert instruction creators, over a dataset of 1,616 linguistic tasks and their natural language instructions, spanning 55 different languages. For both user groups, LINGO promotes the creation of more difficult tasks for pre-trained models, that contain higher linguistic diversity and lower instruction bias. We additionally discuss how the insights learned in developing and evaluating LINGO can aid in the design of future dashboards that aim to minimize the effort involved in prompt creation across multiple domains.
\\

\end{abstract}

\section{Introduction}
\label{sec:intro}

Benchmark datasets play a key role in driving progress in Natural Language Processing (NLP) \cite{rogers2023qa,wang2022benchmarking}. Pre-trained language models (PLMs) have achieved state-of-the- art performance on many benchmark tasks \cite{peters-etal-2018-deep,brown2020language} and have shown promising generalization capabilities \cite{khashabi2020unifiedqa,aghajanyan2021muppet}. However, cross-task generalization, or generalization of a PML to unseen tasks, remains a comparatively hard challenge \cite{mishra2021cross}.

\begin{figure}[H]
    \centering
    \includegraphics[width=0.65\linewidth]{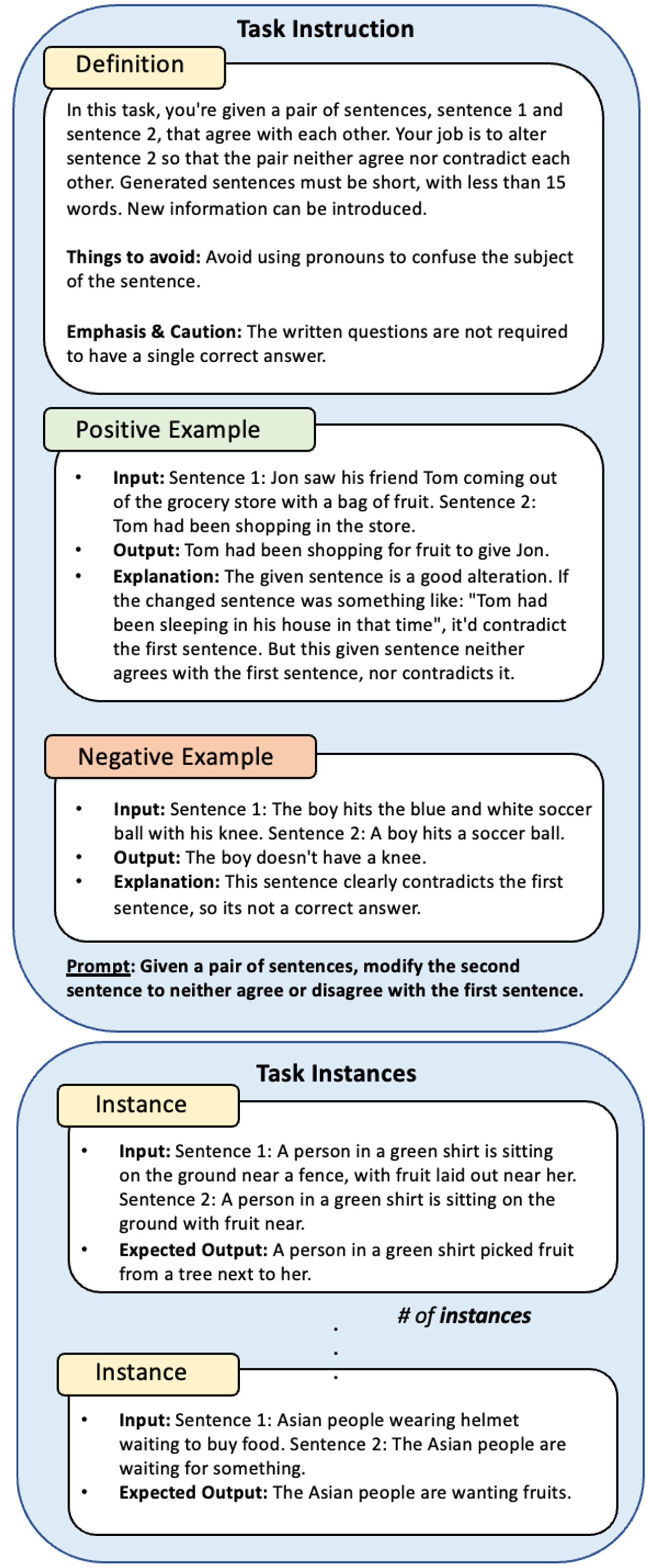}
    \caption{Schema used for representing natural language instructions. The task is created from the SNLI benchmark \cite{bowman2015snli}}
    \label{fig:2}
    \vspace{-1em}
\end{figure}

In contrast, humans are adept at such generalization. For instance, NLP benchmarks are commonly created through crowdsourcing, wherein crowdworkers create instances for a task, following instructions from dataset creators \cite{efrat2020turking,suhr2021crowdsourcing}. Recently, the NLP community has made great progress in building models for generalization to unseen tasks via in-context instructions \cite{brown2020language,chowdhery2022palm}. These instructions comprise a natural language prompt that defines and exemplifies a task for a PLM, similar to crowdsourcing instructions. Figure \ref{fig:2} represents the organization of a natural language instruction (composed of a \textit{task description} and \textit{examples}) for a text modification task, along with task instances for model evaluation.

Unfortunately, prior work in crowdsourcing has shown that task instructions provided by dataset creators often influence crowdworkers to follow specific patterns during instance creation. This leads to collection of biased data, which inflates model performance \cite{geva-etal-2019-modeling,parmar2022don,hettiachchi2021investigating}. This is particularly critical in high-risk domains such as healthcare \cite{Mishra_Arunkumar_2021}, where incorrect answers to a task can prove fatal. Therefore, despite the success of instructions in PLMs, the NLP community is faced with the question of:  \textit{are we truly modeling a task or a user's instructions}? To investigate this, we developed \textit{LINGO}, an end-to-end visual analysis and authoring tool for natural language instructions. LINGO is designed to address real-world challenges faced in NLP benchmark creation, by helping authors identify/compare \textbf{\textit{instruction bias}} pertaining to different types of tasks; LINGO also facilitates real-time task instruction creation/modification and evaluation with PLMs, as shown in Figure \ref{fig:teaser}. \textit{\textbf{To our knowledge, LINGO represents the first visual analytics system that enables realtime feedback and reconciliation opportunities to data creators for instruction bias.}}

In this paper, we describe the process of developing LINGO, based on a pre-study with expert instructional prompt creators to identify salient challenges involved in the identification and reconciliation of instruction bias. We robustly evaluate LINGO via a real-world usage scenario, and a user study with both novice and expert instruction authors over a dataset of 1,616 linguistic tasks and their natural language definitions, spanning 55 different languages. \red{(We note that, while LINGO is language-independent, linguistic knowledge is required for users to analyze multi-lingual tasks}.) The results indicate that LINGO promotes the creation of more difficult tasks for pre-trained models, that are defined with higher linguistic diversity and exhibit lower instruction bias. In following a design study methodology, and based on the process of creating and evaluating LINGO, feedback from our formative interviews and the user study, we additionally discuss how tools like LINGO can benefit the NLP and AI communities in using prompting to create less biased benchmark datasets and evaluate PLMs, by reducing the effort involved in diverse prompt creation across multiple domains and languages, and helping standardize the quantification and elimination of bias in text data.

\begin{table*}[t]
\centering
\scriptsize
\resizebox{0.8\textwidth}{!}{%
\begin{tabular}{@{}lcccc@{}}
\toprule
Meta-Dataset                              & \begin{tabular}[c]{@{}c@{}}$\textsc{Sup-NatInst}^{+,\bigcirc}$\\ \cite{wang2022benchmarking}\end{tabular} & \begin{tabular}[c]{@{}c@{}}$\textsc{NatInst}^\bigcirc$\\ \cite{mishra2021cross}\end{tabular} & \begin{tabular}[c]{@{}c@{}}\textsc{Prompt-Source}\\ \cite{bach2022promptsource}\end{tabular} & \begin{tabular}[c]{@{}c@{}}$\textsc{Flan}^+$\\ \cite{weifinetuned}\end{tabular} \\ \midrule
Number of Tasks                           & 1616                                                                           & 61                                                                         & 176                                                                             & 62                                                                      \\
Number of Instructions                    & 1616                                                                           & 61                                                                         & 2052                                                                            & 620                                                                     \\
Number of Task Categories                 & 76                                                                             & 6                                                                          & 13                                                                              & 12                                                                      \\
Average Task Definition Length (in words) & 56.6                                                                           & 134.4                                                                      & 24.8                                                                            & 8.2                                                                     \\ \bottomrule
\end{tabular}%
}
\caption{Notable open-source meta-datasets of natural language instructions. $\bigcirc$ : indicates the presence of negative examples in the task instruction. $+$ : indicates the presence of non-english tasks.}
\label{tab:1}
\vspace{-2em}
\end{table*}

\section{Related Work}

\subsection{Learning from Instructions}

The success of PLMs \cite{brown2020language,chowdhery2022palm} has powered the development of various `prompting' techniques \cite{liu2023pre} in NLP.  Typically, prompts are extremely short and may not include a complete definition of complex tasks \cite{schick2021few,reynolds2021prompt}. During prompting, a text snippet is added to unlabelled data; this allows the downstream tasks to be represented in the format as pre-training objectives, and requires no new parameters, enabling model learning without having to finetune for each task. Prompt construction is an active research area; we focus on a technique that extends prompting to better define benchmark tasks, called `natural language instructions'. This approach stems from NLP benchmark creation via crowdsourcing, where instructions are given to crowdworkers for sample creation pertaining to a particular task \cite{geva-etal-2019-modeling,efrat2020turking}. Natural language instructions are a versatile means of defining goals; they have previously been studied in applications such as SQL translation \cite{kim2020natural} and interaction within a visual environment \cite{shridhar2020alfred}. Instructions describe tasks in natural language \cite{weller2020learning}, and guide PLMs to generalize to unseen tasks \cite{mishra2021cross,weifinetuned} without requiring task-specific training (see Figure \ref{fig:2}). Meta-datasets (i.e., datasets of datasets) \cite{triantafilloumeta} of human-authored instructions (as shown in Table~\ref{tab:1}) have been used as benchmarks for cross-task generalization in PLMs. 

\subsection{Bias in Instructions}

Prior work has shown that crowdsourced natural language benchmarks exhibit various spurious biases (i.e., unintended correlations between input and output), that lead to overestimation of PLM performance \cite{schwartz2017effect,poliak2018hypothesis,gururangan-etal-2018-annotation,le2020adversarial}. Several techniques have been proposed to handle such bias post-creation, including improving linguistic diversity of samples \cite{yaghoub2020dynamic,larson2019outlier,stasaski2020more} and augmenting data with adversarial samples intended to fool the model \cite{wallace2019trick,kiela2021dynabench,talmor1commonsenseqa}. Similarly, there is evidence that natural language instructions provided by dataset creators during crowdsourcing influences crowdworkers to follow specific patterns during sample creation \cite{geva-etal-2019-modeling,parmar2022don,hettiachchi2021investigating}. These patterns, termed as `instruction bias,' propagate to the dataset and are subsequently over-represented in the collected data. For instance, the crowdsourced DROP dataset~\cite{dua-etal-2019-drop} defines a reading comprehension task where discrete reasoning (e.g., addition, sorting, or counting) must be done based on a paragraph of text. The given instructions contain examples for reference during sample creation; 70\% of these examples begin with \ttfamily
"How many [field goals | yards | points | touchdowns]"
\rmfamily
Subsequently, $\sim$ 62\% of the crowdsourced data samples in both the train and test splits begin with the same pattern of phrasing~\cite{dua-etal-2019-drop}.

\subsection{Visual Analysis of Natural Language Benchmarks}

The majority of prior approaches for handling data bias in NLP benchmarks focus on the adjustment of  hyperparameters for machine learning models, or on artificially rebalancing the biased datasets post-creation \cite{li2019repair,li2018resound,kaushiklearning,gardner2020evaluating,nie2020adversarial}. For instance, users can be encouraged to rephrase highlighted portions of text that are important for a model to make a prediction to promote adversarial sample creation during benchmark construction \cite{wallace2019trick,kiela2021dynabench,talmor1commonsenseqa,attenberg2015beat,vandenhof2019hybrid}. Visual analysis tools that afford NLP leaderboard probing through metric customization have also been proposed as a way to isolate model biases \cite{mishra2021robust}. The visualization of bias detection algorithm results \cite{lavalle2020approach,lavallemethodology} has been leveraged to inform non-expert data creators on what constitutes data bias, although the specific use of visualization for analyzing NLP benchmarks is relatively underdeveloped. Considering the problem of biased data more generally across machine learning and deep learning, related visualization techniques and tools have been developed to analyze dataset biases, vulnerabilities, predictive fairness~\cite{cabrera2019fairvis,ma2019explaining,wang2020visual,chatzimparmpas2020state}, although these tools are designed to address different tasks and goals than LINGO. More specifically, LINGO examines a wide array of linguistic features to identify and resolve instruction bias. Analytical capabilities are balanced with  visualization techniques catering to both expert and non-expert prompt authors. 
\section{Design Process and Goals}
\label{sec:design_reqs}

To help motivate our system design, we surveyed five NLP practitioners (with 4--6 years of experience in investigating data bias and prompt authoring). Our motivating question was: \textit{what are the salient challenges in the identification and reconciliation of instruction bias, where visual analytics can be a key approach for providing insights/solutions?} Table~\ref{tab:2} summarizes several challenges that were explicitly discussed by the practitioners. All participants noted that bias detection in NLP benchmarks is mainly performed as a post-hoc analysis using backend code and algorithms, however there are no standardized practices or metrics for isolating bias. If implemented, bias reconciliation primarily involves either data augmentation or model restriction. Importantly, visualization-based approaches for bias detection and reconciliation in both NLP instructions and datasets are relatively unexplored.

\begin{table}[H]
\centering
\resizebox{\columnwidth}{!}{%
\begin{tabular}{@{}llll@{}}
\toprule
\multicolumn{4}{l}{Challenges for Analyzing Instruction Bias}                        \\ \midrule
(5) & Wide range of instruction features that can cause bias            & * &\textbf{DR1,DR2}\\
(4) & Difficult to juxtapose samples to identify bias sources             & * &\textbf{DR3}\\
(4) & Different instruction types will exhibit different biases             & * &\textbf{DR1,DR5}\\
(3) & PLM performance depends on instruction quality                             & * &\textbf{DR4}\\
(3) & Instruction examples dictate created sample diversity                      & * &\textbf{DR4,DR5}\\
(3) & Instructions optimal to a model may not be optimal for humans &   &\\ \bottomrule
\end{tabular}%
}
\caption{Our survey with NLP experts identified several items as salient challenges for analyzing instruction bias. We note the number of users who referenced each challenge in parentheses. Items addressed by our current tool are noted with an asterisk, and are mapped to \red{derived design requirements.}}
\label{tab:2}
\vspace{-1em}
\end{table}

\begin{figure*}[t]
    \centering
    \includegraphics[width=0.8\textwidth]{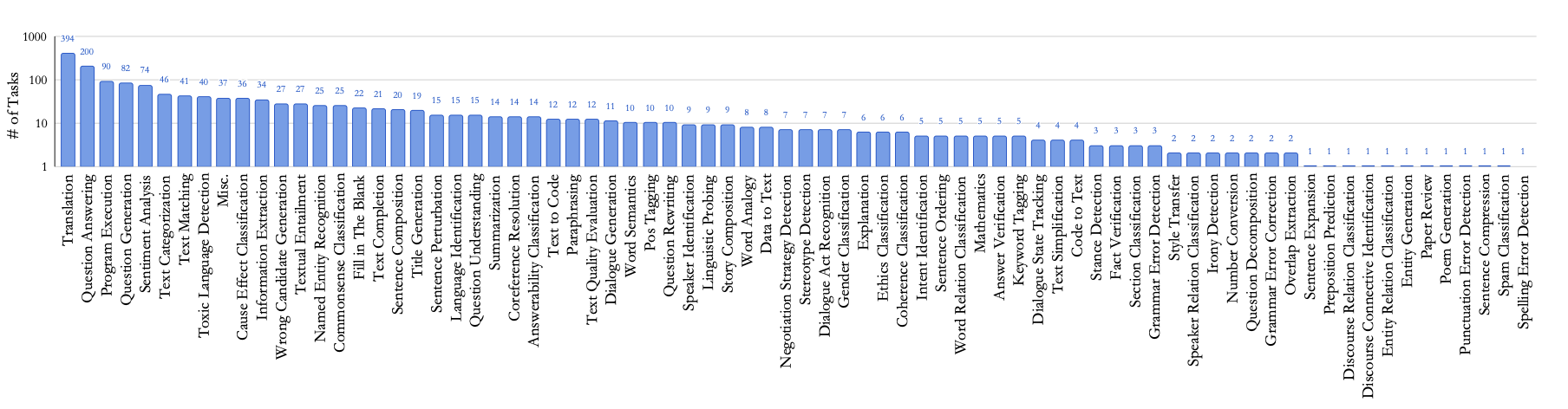}
    \caption{\red{Task categories in \textsc{Sup-NatInst}.} }
    \label{fig:3}
    \vspace{-2em}
\end{figure*}

Based on the challenges listed in Table \ref{tab:2}, we derived a set of five high-level design requirements (\textbf{DRs}) that a visual analytics system can employ for analyzing and reconciling instruction bias. 

\textbf{DR1: Compare instructions in a categorized manner.} Writing style is dictated by the task category, domain (reasoning type), and dataset for which the instruction is created; this must be accounted for when comparing instructions within or between categories.

\textbf{DR2: Allow granular analysis of different parts of an instruction.} Instructions consist of the task definition and examples (positive and negative) with explanation. We consider the entire instruction for analysis as an overview comparison against constituent data samples used for PLM evaluation. However the biases exhibited within and between different parts of the instruction may vary, so multi-granular analysis can be particularly beneficial.

\textbf{DR3: Support juxtaposition of instruction examples and data samples.} The modification of instructions to reduce instruction bias is contingent on the relationship between instruction examples and data samples. Examples must have sufficient inductive bias to inform a PLM/crowdworker of the task, but must also exhibit sufficient diversity so sample creation isn't artificially constrained.

\textbf{DR4: Evaluate PLM on modified instructions in situ.} Instruction effectiveness must be gauged in situ in terms of PLM performance. This ensures that instruction modification to reduce instruction bias still retains sufficient inductive bias for the model to learn.

\textbf{DR5: Provide multiple perspectives for analysing bias features.} Instruction bias can arise from a wide variety of linguistic features. Task-agnostic features must be selectable to facilitate analysis of different instruction categories. 

Notably, a significant constraint in successfully addressing \textbf{DR1--DR5} is that the target user base (prompt authors) likely lacks visualization expertise. Therefore, in keeping with a design study methodology \cite{sedlmair2012design}, when implementing LINGO we adhered to the standard of a ``well-justified combination of [primarily] existing techniques.'' This allowed us to balance the analytical insight and power provided by LINGO, while not overwhelming them with overly complex or esoteric visual encodings, or requiring a steep learning curve. We also note that LINGO is designed to potentially benefit both expert and non-expert prompt authors during instruction creation for PLM evaluation; additionally, instructions for crowdsourced data collection can be improved, based on insights gleaned by dataset creators. 


\section{Dataset Schema}

To demonstrate LINGO, we use a state-of-the-art meta-dataset of task instructions, called Super-Natural Instructions (\textsc{Sup-NatInst}) \cite{wang2022benchmarking}. This benchmark consists of 1,616 NLP tasks along with their natural language instructions; 76 task categories span 55 different languages (576 non-english tasks) are present (Figure \ref{fig:3} shows the diversity and sizes of the 76 task categories). These tasks were contributed by 88 NLP practitioners in a crowdsourced manner. \textsc{Sup-NatInst} is considered state of the art for model evaluation on task instructions.

Instructions are represented via a unified schema, as shown in Figure \ref{fig:2}. Broadly, schema ingredients are defined as follows:

\begin{itemize}
    \item \textbf{Definition} defines a given task in natural language. This is a complete definition of how an input text (e.g., a sentence or a document) is expected to be mapped to an output text.
    \item \textbf{Positive Examples} are samples of inputs and their \textit{correct} outputs, along with a short explanation for each.
    \item \textbf{Negative Examples} are samples of inputs and their \textit{incorrect} or \textit{invalid} outputs, along with a short explanation for each.
\end{itemize}

Task instances (i.e., data samples to be solved by the model given the instruction) are organized as textual input and a list of acceptable textual outputs. The number of instances is limited to 6.5K per task to avoid instance imbalance, as done for model evaluation in \textsc{Sup-NatInst} \cite{wang2022benchmarking}. 

A single dataset can be used to generate multiple subtasks; for instance, the SNLI dataset used for the task in Figure~\ref{fig:2} is used to generate 12 tasks in total for \textsc{Sup-NatInst} pertaining to the task categories of answer generation, wrong answer generation, text modification, and classification.


\begin{table}[H]
\centering
\resizebox{0.9\linewidth}{!}{%
\begin{tabular}{p{0.25\linewidth}p{0.75\linewidth}}
\toprule
Bias Category & Features Evaluated\\ \midrule
Diversity     & Sample Length \cite{wallace2019trick}, Unique Vocabulary (new words introduced) \cite{yaghoub2020dynamic,larson2020iterative} \\
Similarity    &  N-gram/POS-tag (part-of-speech tags: nouns, verbs, adjectives, adverbs): Frequency, Overlap, Jaccard Similarity \cite{wallace2019trick,yaghoub2020dynamic,larson2019outlier,poliak2018hypothesis,gururangan-etal-2018-annotation}\\
Component Bias$^\triangle$    &  \textsc[ENT] $\times$ \textsc[ENT] Overlap \cite{larson2019outlier}, \textsc[ENT] $\times$ \textsc[ENT] Correlation \cite{stasaski2020more,wallace2019trick} \\ \bottomrule
\end{tabular}%
}
\caption{LINGO supports three bias categories, each composed of several available metrics. $^\triangle$ \textsc[ENT] represents \ttfamily\{task instruction | definition | example (positive/example/both) | task instance\}\rmfamily}
\label{tab:3}
\vspace{-2em}
\end{table}

\section{Instruction Bias Measures}
\label{sec:5}

LINGO supports three broad categories of instruction bias measures, based on existing bias identification measures for NLP benchmark analysis. Table~\ref{tab:3} defines the categories of instruction bias supported in the system, along with the evaluation metrics curated from prior research for each category. (LINGO is also extensible to accommodate additional measures.)

Broadly, \textbf{diversity} examines an individual task instruction, while \textbf{similarity} juxtaposes multiple task instructions; both utilize the full instruction in calculation. \textbf{Component bias} compares different parts of instructions and task instances; this can be done (i) within or between tasks (e.g., a positive example from Task X can be compared to a task instance either from Task X or any other task), and (ii) at multiple granularities (sentence/word/part-of-speech tag/n-gram). Metrics are calculated using the NLTK library~\cite{loper2002nltk} with a Python API; text pre-processing involves stop-word removal, tokenization, POS tagging, and lemmatization.

\section{LINGO}

We now describe LINGO’s system design. The system is a full-stack application, with a backend server for data storage, query, and NLP-based computation, and  a frontend interface for visualization and interaction. For examples of how LINGO can be used to analyze and reduce instruction bias, see the usage scenario in \S\ref{sec:7} and the demo video include in the supplemental materials.

\subsection{Backend}
\label{sec:backend}

The backend server is built using Python and Flask, and acts as the storage and service layer between the instruction dataset and the frontend user interface. A MongoDB \cite{MONGODB} database stores the tasks as JSON objects, supporting quick retrieval.

Based on a user's choice of (i)~task instruction(s) to analyze, (ii)~components of the instruction to include in the analysis, and (iii)~bias metric(s) to be used, we retrieve instruction(s) from the database (this supports \textbf{DR2}) and perform necessary metric computations (see \S \ref{sec:5}) to populate the front-end panels (thus supporting \textbf{DR5}). We evaluate a subset of task instances using a Python script to call on the GPT-3 \cite{brown2020language} Open API (davinci-instruct-beta-v3), on load/modification of task instructions (\textbf{DR4}). We follow \cite{mishra2021cross} and treat the tasks as text-generation problems, using ROUGE-L \cite{rouge2004package} to automatically evaluate the generated outputs and visualize the results.

\begin{figure*}
    \centering
    \includegraphics[width=0.7\textwidth]{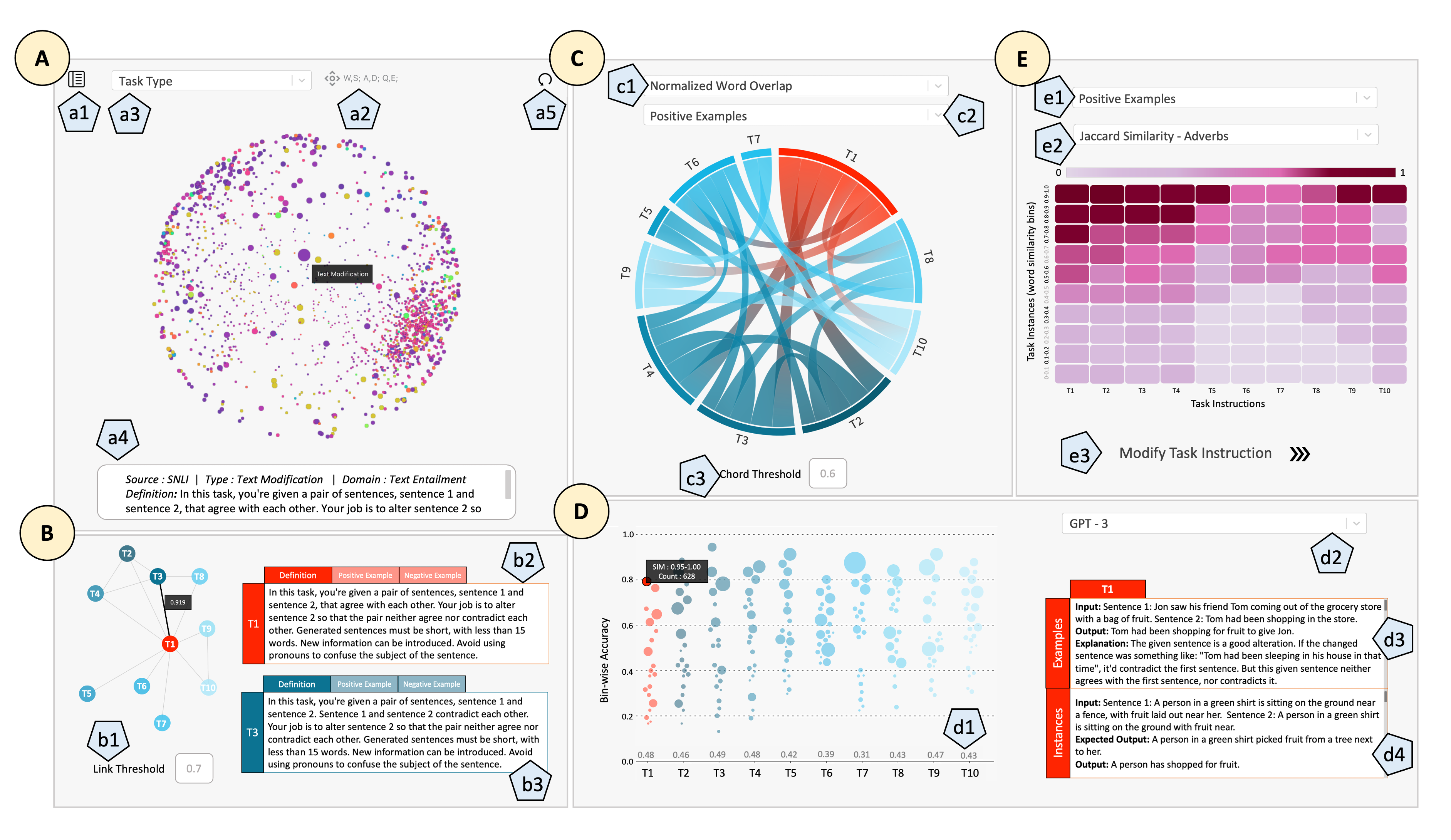}
    \caption{LINGO consists of 5 panels to support (A) overview and selection of instructions, (B) examination of instruction correlation, (C) instruction decomposition and comparison, (D) model evaluation with instructions, and (E) inspection and resolution of instruction bias.}
    \label{fig:4}
    \vspace{-2em}
\end{figure*}

\subsection{Frontend}

Figure~\ref{fig:4} shows an overview of LINGO’s frontend interface. It consists of five panels \textbf{(A)}–-\textbf{(E)} which are linked via coordinated interactions (e.g., hovering on or selecting a task instruction in one view highlights or selects the instruction in other views) and designed to support the five \textbf{DRs} described in \S\ref{sec:5}. \red{We use fixed color scales that accommodate users with color-impaired vision using the R colorspace toolbox.} 

\textbf{(A) Overview Panel.} The overview panel consists of a t-SNE projection, where each point is a task instruction. The pairwise distances between instructions is based on sentence embedding distances, calculated over the complete task instruction (i.e., \ttfamily definition+examples\rmfamily). \red{We use t-SNE here as it ably preserves geometry for nested clusters of varying density, though LINGO is easily extensible to support MDS, PCA, etc.}
This projection can be rendered both on a traditional 2D plane, or as a rotatable 3D sphere; \red{we model the embedding sphere after embedding projectors used in TensorFlow}. Regarding this latter encoding option, in particular we found during early testing that a 2D projection was at times overly cluttered, which made it difficult to identify and select individual points (in particular, translation tasks, which accounted for $\sim$24\% of the tasks in the dataset used, often contained very similar phrasing). Toggling to a 3D projection could provide a better separation of points and identification of clusters. \red{However, filtering is sometimes necessary to enhance scalability and prevent clutter for larger or more closely distributed datasets.} Point color represents the categorization of the instructions based on task type/domain/source dataset \textbf{(DR1)}. \textbf{(a1)} A color legend can be viewed by clicking on a toggle icon. \textbf{(a2)} allows interacting with the sphere projection; rotation along x/y/z axes is done via arrow keys. \textbf{(a3)} The basis for categorizing task instructions can be changed with the dropdown; the default categorization is based on task type (i.e., the nature of the task). Mousing over of a point highlights all task instructions belonging to the same category; a tooltip details the category of the task and \textbf{(a4)} a text area shows the source dataset, domain, and definition of the task; \red{users can also search for a specific task by typing in this text area}.  Clicking on a point selects the corresponding ``root instruction'' for further analysis, along with the $k$-most highly correlated instructions with respect to the root instruction ($k$ default: 9). For the selection, the root node is colored red, and the additional $k$ nodes are assigned blue shades based on their correlation amount.  
Additional instructions in the same category as the root instruction remain colored with decreased opacity; all other points are toggled to grayscale with decreased opacity. \textbf{(a5)} Clicking a refresh icon clears the selection, resets the sphere to the default view, and clears other panels in the display.

\textbf{(B) Correlation Panel.} This panel displays the selected root and $k$ instructions (from \textbf{(A)}), with a node-link diagram. Each node represents an instruction, colored and labelled in order of correlation (sentence similarity). (T1 represents the root instruction and is colored red, T2 represents the most correlated node, and so on.) 
Links between nodes indicate a similarity score over a threshold (adjustable using \textbf{(b1)}, default: 0.5). Mousing over (or clicking) a node displays its similarity score in a tooltip, as well as the text of the corresponding root and task instructions (in \textbf{(b2)} and \textbf{(b3)}). Mousing over (or clicking) a link displays the instructions corresponding to the end nodes in \textbf{(b2)} and \textbf{(b3)} and also the similarity between the nodes in a tooltip. The tabs in \textbf{(b2)}/\textbf{(b3)} allow viewing either the full instruction/definition/positive/negative examples for comparison \textbf{(DR2)}. 
\red{We preserve positional information from the 3D embedding space through the use of these network links, which indicate the strength of the relationship between two tasks at different granularities. The color scale assigns more similar samples with darker colors. Also, the similarity thresholds to draw links and the k-value can be customized by the user; therefore the range of similarity considered varies. (B) hence streamlines the selection of a subset of tasks from A that implicitly provide contextual information for the root task instruction. This is particularly important when using models such as ChatGPT which can remember several previously entered instruction prompts.} 

\textbf{(C) Instruction Decomposition Panel.} This panel shows inter-task relationships between the selected root and $k$ instructions using a chord diagram. Multi-granular comparison can be done, based on sub-components of the task instruction chosen for examination. Specifically, the inter-task relationship is examined in terms of normalized length/normalized word overlap correlation (chosen from \textbf{(c1)}). Sub-components that can be used for analysis are: \ttfamily\{task instruction | definition only | example only (positive/example/both)\}\rmfamily \textbf{(DR2)}, selected using \textbf{(c2)}. The task labels and color scheme follow (B); chord ribbons are drawn between tasks with relational values over a threshold (see \textbf{(c3)}, default: 0.5). Color corresponds to task color, and thickness represents the strength of the relation. Mousing over (or clicking) a chord highlights its corresponding tasks with labels displaying the similarity values. Additionally, the corresponding tasks are highlighted in (\textbf{B)}, and \textbf{(b2)} and \textbf{(b3)} are populated. \red{(C) explicitly highlights which sub-components of the task instruction that contribute the highest bias. By tracking the linguistic diversity used for low-bias tasks connected to the root instruction with chords, users can understand the scale and nature of instruction modification required.}

\textbf{(D) Model Results Panel.}
This panel displays the model evaluation results for task instances corresponding to the subset of instructions in (B); \textbf{(d2)} is used to select the model for evaluation. As noted in \S\ref{sec:backend}, we use GPT-3 \textbf{(DR4)}. Each beeswarm plot represents a task (T1--T10); ordered and colored by its correlation to the root instruction. The y-axis plots the accuracy of GPT-3 on the task instances. Each task instruction is evaluated by testing the model on 6.5K task instances; \textbf{(d1)} denotes the GPT-3 accuracies for all tasks. To reduce clutter, we perform an aggregation step: the task instances are grouped into 20 equidistant bins based on their word similarity to the positive and negative examples given in the task instruction as $[0.00-0.05],[0.05-0.10],...,[0.95-1.00]$ \textbf{(DR3)}. The average accuracy over instances belonging to each bin is plotted as the y-axis value (meaning each beeswarm displays up to 20 points). The size of the point represents the number of task instances present in that bin  (if a bin is empty, it is not plotted as a point). Mousing over (or clicking) a points displays: (i)~a tooltip with the number of instances, and similarity bin limits, (ii) a set of positive and negative examples for the corresponding task in \textbf{(d3)}, and (iii) a set of three task instances from the corresponding similarity bin, which have the highest word overlap with the examples in \textbf{(d4)}. \red{To summarize, this panel examines the influence of instance-example similarity bias on model performance, i.e., are the examples given in the task instruction sufficiently diverse to enable to model to generalize to unseen instances?}

\begin{figure*}[t]
    \centering
    \includegraphics[width=0.85\textwidth]{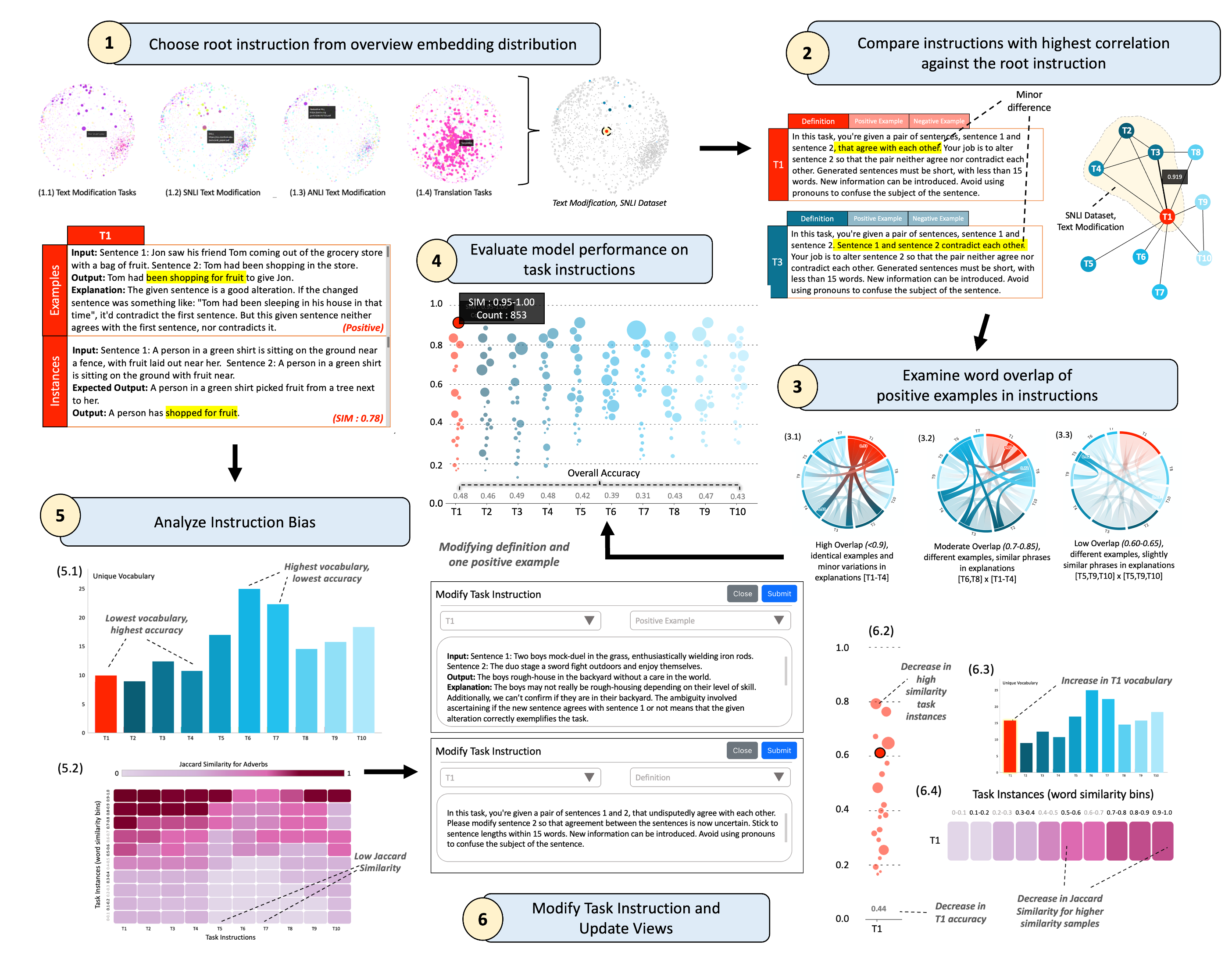}
    \caption{The interactions taken during the Usage Scenario described in \S\ref{sec:7}.}
    \label{fig:5}
    \vspace{-2em}
\end{figure*}

\textbf{(E) Bias Metrics Panel.} This final panel can be used to asses instruction bias using metrics that gauge diversity and similarity of instructions (see Table~\ref{tab:3}). \textbf{(e1)} The user can can select the sub-components of the instruction to analyze (DR2), \textbf{(e2)} and can also select desired bias metrics \textbf{(DR5)}. Frequency based metrics (for n-grams and POS Tags), sample length, and unique vocabulary count are represented using bar charts, where each bar corresponds to a task (Figure~\ref{fig:5}\textbf{(6.3)} shows an example). These metrics are only applied to the task instruction, and measure the diversity of the task. Overlap (normalized) and Jaccard Similarity for n-grams and POS Tags is plotted with a heatmap, where increasingly dark pink shades represent values closer to 1. These metrics compare the similarity of the task instruction to the task instances. The task instances are grouped into 10 equal-spaced bins as $[0.0-0.1]...[0.9-1.0]$, based on normalized word similarity to instruction examples). Clicking \textbf{(e3)} opens a dialog window where the definitions and/or examples of any one task instruction (from T1-T10) can be modified (see demo video,  Figure~\ref{fig:5}\textbf{(6.1)}). On submitting the new instruction, panels \textbf{(A)}--\textbf{(E)} update accordingly, where the modified task instruction now serves as the new \textit{root instruction}. \red{We use either heatmaps or bar charts to represent cross-task bias statistics. This is done to differentiate the nature of the quantities being compared, i.e., as normalized ratios or absolute counts. Thus, users are less likely to conflate bias-metrics used to identify which specific modification patterns will optimally reduce instruction bias.}

\section{Usage Scenario}
\label{sec:7}

To help illustrate how LINGO can be used to analyze instruction bias, we now present a use case scenario with Owen, a prompt author who is analyzing instructions created for Text Modification (TM) tasks \red{(see supplementary material for additional case studies)}. Figure \ref{fig:5} shows his workflow.

\textbf{(1)} Owen first searches for instructions corresponding to TM using the overview panel. He notices that TM instructions show large embedding distances \textbf{(1.1)}, with the formation of local clusters if instructions use the same source dataset or domain \textbf{(1.2, 1.3)}. This is in contrast to other tasks like translation, where a majority of points are clustered in the same region across source datasets \textbf{(1.4)}. Owen selects a TM root instruction, originally from the \textsc{SNLI} source dataset \cite{bowman2015snli} for further analysis. The root instruction is colored red (T1), and the 9 most highly correlated instructions (T2-T10) are colored in shades of blue.

\textbf{(2)} In the Correlation View, Owen sets the threshold for links to $0.7$. He notes the variation in correlated instructions' task domains, types, and sources (see Table \ref{tab:4}). On further examination, Owen realizes that all the TM instructions use the same SNLI source dataset as T1. Another interesting instruction is T8. It has links to T1 and T3, and belongs to the same task domain: Text Entailment. T2--T4 use nearly verbatim definitions in task instructions when compared mutually, as well as against T1, as highlighted in Figure \ref{fig:5}. 

\textbf{(3)} Owen decides to compare the word overlap between the positive examples of T1--T10 (chord threshold: 0.6), in order to understand why different task types are highly correlated. He finds that between T1--T4, positive examples show high word overlap \textbf{(3.1)}; on examining the text, they are seen to only marginally differ in explanation. T6 and T8 use similar explanation phrasing to T1--T4 for positive examples \textbf{(3.2)}. Similarly, word overlaps between the T5, T9 and T10 examples arise in the explanations of the positive examples \textbf{(3.3)}.

\begin{table}[h]
\centering
\resizebox{0.9\linewidth}{!}{%
\begin{tabular}{@{}lllll@{}}
\toprule
Task ID & Source Dataset & Task Type & Task Domain   & Correlation $[0,1]$\\ \midrule
\textbf{T1} & \textbf{\textsc{SNLI}} & \textbf{Text Modification} & \textbf{Text Entailment} & \textbf{1.000}\\
T2      & \textsc{SNLI}           & Text Modification   & Text Entailment & 0.920\\
T3      & \textsc{SNLI}           & Text Modification   & Text Entailment & 0.919\\
T4      & \textsc{SNLI}           & Text Modification   & Text Entailment & 0.911\\
T5      & \textsc{ROC Stories}    & Sentence Ordering   & Deductive       & 0.858\\
T6      & \textsc{XCOPA}$^+$      & Classification      & Causal         & 0.850 \\
T7      & \textsc{ASSET}          & Text Simplification & Commonsense      & 0.842\\
T8      & \textsc{SherLIiC}       & Text Entailment     & Text Entailment  & 0.838\\
T9      & \textsc{ROC Stories}    & Text Completion     & Commonsense     & 0.836 \\
T10     & \textsc{ROC Stories}    & Text Completion     & Commonsense     & 0.836 \\ \bottomrule
\end{tabular}%
}
\caption{Tasks with most highly correlated instructions, with respect to a selected root instruction (T1). $+$ : non-english input.}
\label{tab:4}
\end{table}

\textbf{(4)} On observing the GPT-3 results for task instances from T1--T10, Owen notes a general trend across all the datasets: task instances in bins with higher word similarity ($>0.7$) to the instruction examples are more accurately solved. Owen confirms that T1--T4 and T7 show similar performance distributions, with most of the high-performing task instances being in high similarity bins. T5 and T8--T10 have a greater proportion of task instances located in bins ranging from [0.55-0.75]; these bins achieve $\sim$20-35\% accuracy. T6 is an outlier with a more uniform distribution of task instances in similarity bins, all ranging from $\sim$20-30\% accuracy. Owen partially attributes this to T6 containing non-English input. Owen then juxtaposes top ranked task instances belonging to bins [0.85-0.90] against instruction examples for T1-T4. He notes that these task instances follow similar topics to the examples, and are solved using similar output patterns, as highlighted in Figure \ref{fig:5}\textbf{(5)}.

\textbf{(5)} Owen checks the unique vocabulary contributed by examples in each task instruction \textbf{(5.1)}. He finds that T6 and T7 contribute the highest unique vocabulary, while T1--T4 contribute the lowest. T6 and T7 are also the hardest for the model to solve (as previously seen in \textbf{(4)}). Next, Owen examines the Jaccard Similarity of adverbs across full task instructions compared against their respective task instances \textbf{(5.2)}. Here, T5--T7 show the lowest values, which indicates higher variation in the data. Owen can partially attribute this to non-repetition of phrases from the definition in the example explanations. 

\textbf{(6)} Finally, Owen modifies the task instruction for T1, in order to (i) reduce the similarity bias between instruction examples and task instances, and (ii) reduce overlap between the instruction definition and instruction example explanations. He updates the definition and replaces a positive example, so that the instruction now contributes a higher proportion of vocabulary and adverbs \textbf{(6.1)}. Owen notices that the beeswarm updates \textbf{(6.2)} for T1 to show decreased overall performance, with  fewer samples in similarity bins $>0.9$ as well as fewer samples crossing 80\% accuracy. Additionally, T1 now exhibits lower Jaccard similarity \textbf{(6.3)} and higher unique vocabulary \textbf{(6.4)}. \red{Overall, Owen's modifications achieve a decrease  in accuracy, Jaccard similarity for words, POS-tags, and n-grams, and an increase in unique vocabulary}. Hence by iteratively changing the task instruction (for instance, replacing more examples) Owen can further reduce instruction bias for T1, and create a more difficult task instruction for the model to solve.

\section{User Study}
\label{sec:user}

To further evaluate LINGO from an empirical perspective, we conducted an in-person qualitative usability study composed of both novice and expert prompt instruction creators, using the \textsc{Sup-NatInst} dataset. The study utilized think-aloud protocol, as well as a post-study questionnaire and feedback session about the experience of participants using LINGO. This allowed us to robustly understand how the insights afforded by LINGO (as well as its overall usability) compared both for novice users and domain experts.

Participants began the study by completing a short survey to collect demographic information and background knowledge. After this, three stages were run:

\textbf{Training.} The participants were first given an overview and usage scenario (with text modification tasks) of LINGO's functionality, and could ask questions and play with the system until they feel confident to proceed. \red{Study participants were familiarized with using the visual idioms, interactions, and color mappings (since explicit legends are not provided in the interface).}

\textbf{Exploration.} Participants were then allowed to freely explore the system for up to 30 minutes. To constrain the study design, we limited participants to analysis of question-answering task instructions. During this stage, participants use think aloud protocol to verbalize their thought processes and actions \cite{fonteyn1993description}.

\textbf{Review.} Participants then completed a short feedback survey about the system's affordances and user experience using Likert scale ratings (1 – strongly disagree, 7 – strongly agree). Participants also had the opportunity to provide freeform comments, suggestions, and criticisms about the interface and their experience.

\textbf{Participants.} We recruit 16 participants (age $\mu$ = 25.46 year, $\sigma$ = 3.04, 10 males and 6 females) studying Computer Science from <Anonymous University>. All participants were proficient in English and had normal (corrected if necessary) eyesight. Participants reported familiarity with instruction authoring based on a 7-point Likert scale: $\mu$ = 4.26 ($\sigma$ = 2.14). All study participants reported relative non-expertise with data visualization (familiarity: $\mu$ = 2.33, $\sigma$ = 1.68).  LINGO was displayed in Google Chrome on a 24-inch monitor ($3840\times2160$ resolution) with keyboard and mouse, connected to a MacBook Pro running macOS Monterey. QuickTime Player recorded session screencasts and audio. We classified users as novices (familiarity: 1-3, 7 users: n1--n7, $\mu$ = 2.38, $\sigma$ = 0.61) or experts (familiarity: 4-7, 9 users: e1--e9, , $\mu$ = 5.41, $\sigma$ = 1.72) for the purpose of post hoc analysis. None of the novice users had previously taken a graduate course in NLP or authored instructions; two had previously participated in NLP benchmark crowdsourcing. On average, novice users spent 19.37 minutes ($\sigma$ = 3.44 minutes) and expert users spent 25.21 minutes ($\sigma$ = 1.92 minutes) using LINGO.

\subsection{System Ratings from Survey Response}

\begin{figure}
    \centering
    \includegraphics[width=0.9\linewidth]{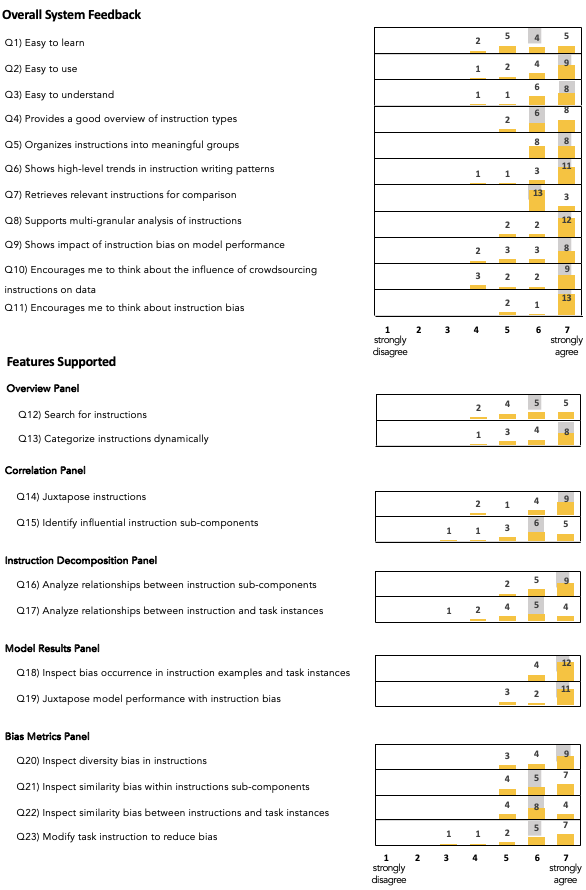}
    \caption{Overall ratings from the post-study survey about various system aspects. Median ratings are indicated in grey.}
    \label{fig:6}
    \vspace{-2em}
\end{figure}

Figure \ref{fig:6} summarizes study results from the quantitative survey. (Note: All results are found to have $p \leq 0.02$ based on running a t-test.) This figure aggregates both novice and expert responses, as general trends remained consistent for both user groups. Overall system ratings (Q1–-Q11) were generally positive, including that it is easy to learn, use, and comprehend (Q10–Q3), it supports meaningful analysis (Q5-–Q9), and encourages participants to think about instruction bias during prompt authoring and crowdsourcing (Q10–-Q11). Responses about specific interfaces features (Q12–-Q21) indicate that they are also generally well received.

\subsection{Support for User Insights}
\label{sec:9.2}

Based on the recorded think-aloud utterances by our seven novice and expert participants (n1--n7 and e1--9), as well as freeform feedback collected during the review stage, we performed an open coding technique based on grounded theory~\cite{walker2006grounded} to understand the types of support, insights, and user experiences that LINGO specifically promoted to our study participants. We summarize four key takeaways below.

\textbf{For novices, the interface was paradoxically both easy to use and too complex.} 
We received several comments from novice participants about LINGO being \textit{``easy to use''}(n6,n5) and \textit{``cohesive and easily navigable pipeline''}(n1,n7). \textit{``I like how you can easily focus on different sub-components of the instruction and that changes how its relationships look... [it] helps me streamline my analysis''} (n7). Paradoxically, other users sometimes felt overwhelmed by the system's available features, such as the number of bias metric choices that were available. \textit{``I'm keeping up until the model results, it's difficult to juxtapose them with the bias metrics" } (n3). \textit{``The chord diagram was new to me, though I got the hang of it after a couple of minutes''} (n1). Multiple users thought that follow-up sessions or helper widgets would lead to more efficient analysis: \textit{``I'd like some more time to go through the tasks, so I understand how to use the bias metrics to change the instruction''} (n3,n2). \textit{``[S]omething like a widget that tells you in plain text what the high-level impact of this bias metric is... helps clarify how to translate bias identification to instruction modification"} (n3,n4). This mixed feedback is likely due to a lack of domain knowledge by some novices, which hampered their user experience.

\textbf{Novices echoed experts when defining instruction bias post-study.} 
Many of the novice and expert participants agreed in conjecturing that instruction bias likely arises when: (i) \textit{``the examples are all too similar amongst themselves"} (n1,n2,e4), (ii) \textit{``the example explanations reiterate the definition exactly, without new contextual information"} (n3,e9), (iii) \textit{``the beeswarm is top-heavy, meaning lots of high similarity instances exist and are correctly solved"} (n7,e2), and (iv) \textit{``if one person creates multiple instructions for a dataset, they are probably following the same patterns throughout...they use the same words, phrases, and topics for examples"} (n3, e8). LINGO is therefore able to support effective and accurate analysis by both novel and expert users. 

\textbf{The beeswarms were effective for recognizing how instruction bias inflates model performance.}
Several participants appreciate the beeswarm representation of task instances as being \textit{``... an intuitive way to represent a large amount of data"}(e1,e3,n2,n5). \textit{``I like that you can drill down and see representative task instances... compare them with the examples"}(n4,e6). Participants also found it useful to use the beeswarm and chord diagrams to filter out which bias metric and tasks to focus on. \textit{``I decided that these [four tasks] have a lot of mutual connections and the beeswarms are top-heavy, so I'm going to check the Jaccard similarities for those tasks"} (e2). \textit{``See how Task 6 has low accuracy but Task 3 has high accuracy? Let's check the unique vocabulary contribution... I was right [Task 6] has it higher} (e8). \textit{``If you look at sentence lengths, you can see longer instructions with shorter example explanations seem to have better performance"} (n1). Several experts also found in situ model results useful while modifying instructions--\textit{``... [results] ground how I'm changing the instruction. I can see what does and does not work to reduce instruction bias, and further ensure I still have sufficient inductive bias for the model."} (e2) and \textit{``I can see this being applicable to analyze bias any sort of benchmark data. We have the human, model, and metrics all simultaneously involved"} (e4).

\textbf{Participants are able to identify potential caveats to reducing instruction bias.}
Several participants pointed out that reducing instruction bias requires a learning curve in identifying what constitutes a positive change to reduce model performance. \textit{``I thought putting lots of synonyms in would help...I guess I need to restructure the definition completely."} (n4,n5). However, participants also recognized that the nature of changes might lead to unnatural language. \textit{``If I keep paraphrasing [this way] to reduce model performance, I've got a shortcut to doing the bias reduction... but the instruction looks strange now so this wouldn't happen in the real-world"} (n1,e2). A couple of participants also questioned the overall effect of the changed instruction on dataset bias: \textit{``I could be making a only a local optima... my highest correlated instruction group will change."} (e3,n4).
\vspace{-2em}
\\
\section{Discussion}


Based on our conducted usage scenario (\S\ref{sec:7}) and user study (\S\ref{sec:user}), LINGO robustly supports \textbf{DR1}--\textbf{DR5} described in \S\ref{sec:design_reqs}. Here, we discuss several takeaways from the design, implementation, and evaluation of LINGO, to synthesize generalizable actionable insights for visualization and NLP communities.

\textbf{Novice users recognize broad sources of instruction bias; experts can see research possibilities.}
While both of our user study groups could use LINGO to analyze task instructions, one significant takeaway from the study was that the types of insights differed based on the user’s experience. Novice users could identify instances were instruction bias exists, but generally could only make broad conclusions about bias. Experts, already familiar with instruction bias and its impact, could leverage LINGO to explore deeper nuances about bias. Subsequently, their instruction modifications were generally more effective in reducing bias. As the pre-study in \S\ref{sec:design_reqs} did not explicitly consider various levels of user expertise, such a result is not wholly unexpected. Future analysis systems for novices can be tailored to accommodate their lack of domain expertise.

\textbf{Explicitly flagging bias in-text while authoring.}
Several novices noted that instruction modification proved challenging, given the potential that a modification could introduce new types of instruction bias; several requested that bias be explicitly highlighted in text. While this is possible, there is a risk that ex ante labeling could render such users dependent on the classification outputs of a model or annotator, again constraining the patterns they might follow during instruction authoring and data creation. A possible strategy might be to allow users to define and highlight patterns of interest during analysis, that will be subsequently flagged when they modify an instruction, but we leave this as future work.

\textbf{The need for visualization tools in NLP}
Expert participants in our study were long-time researchers ($+$ 4 years) in NLP, but \textit{none had previously encountered or used visual analytics tools focusing on NLP text bias}. While many backend algorithms and text-highlighting driven interfaces catering to bias identification and resolution have previously been published in the NLP community (e.g., \cite{kaushiklearning}), there is a lack of standardization and adoption. In view of this, tools like LINGO can simplify the overhead involved in bias identification post-creation, by providing an in situ pipeline to procedurally target bias during text authoring. 

\textbf{Designing for non-expert visualization users.}
Even though our user study participants were non-experts in visualization, they were still able to use the system to probe task instructions and glean insights to reduce instruction bias.  LINGO's focus on selecting and utilizing familiar visual techniques therefore provides a cohesive pipeline for instruction analysis, that facilitates a lower learning curve. \red{Particularly, the provision of visually distinct panels to analyze different bias aspects reduces the potential for users to conflate/become overwhelmed by different choices for bias analysis, e.g.: users shortlist which aspect of bias to drill down on for instruction modification, despite experiencing visual novelty, by using panels in tandem.}
LINGO further motivates the necessity to standardize bias quantification and elimination in NLP. This is a non-trivial problem, but we believe visualization can likely provide an important step in the process, by helping researchers identify where the principal issues are present.

\textbf{System Limitations and Future Work}
In particular, the user study helped us to identify several areas where LINGO (or tools like LINGO) could be improved in or expanded to the future. In particular, we note three here:


\textit{(1) Recommendations for Instruction Modification:}
Instruction modification was at times difficult for novice authors, given that instructions must retain sufficient inductive bias for the model to learn after modification. Particularly, for tasks like custom text generation, specialized contextual information must be provided in the instruction. For instance, in order to create instructions for a model to write the bio of a user, the user must provide sufficient personal information. The authoring/modification of such task instructions is therefore a complex process. Approaches like \cite{Mishra2022HELPMT} involve prompting models to ask relevant questions to reduce the cognitive load on the user, in identifying key information specific to a particular task type. Such approaches can be extrapolated to provide recommendations during instruction modification that are tailored to reducing the level of bias present in an instruction. \red{Such bias reduction techniques could therefore be less likely to introduce new types of bias}.  Another relevant approach to author instructions with richer contextual information is to provide the language model with multiple variants of the same instruction at once, which define the task from different perspectives \cite{puri2022many}. This has proven useful for improving model performance, particularly on low-data tasks.

\textit{(2) Expanding Evaluation:}
While LINGO successfully incorporates human, model, and metric-based feedback in our analysis paradigm, the model and metric-based feedback is restricted by the chosen domain. Instruction bias examination necessitates the use of PLMs capable of accepting instructions as input, along with a limited number of task-agnostic metrics to gauge bias. Future studies can broaden the scope of LINGO's analysis to quantitatively identify textual data bias at scale, for both instructions and NLP benchmarks, across a wider array of language models and with customized bias metrics \cite{Mishra2020OurEM,Mishra_Arunkumar_2021}; \red{LINGO is easily extensible to integrating new metrics in either the front/backend to better support expert users}. By quantitatively evaluating ``data quality'' in this manner, it may be possible to establish more standardized procedures for bias quantification and resolution, leveraging visualization for analysis and verification. 

\textit{(3) Extensibility to Prompting for Multimedia Data}
LINGO focuses on text-based task instructions, covering diverse applications, that do not involve images and are purely based on text data. A key challenge in extending LINGO to analyze prompts for multimedia data is the potential lack of human-interpretable bias features. For instance, ``concepts'' (a group of pixels that express a meaningful notion in an image) are an explanation technique for image classification~\cite{kim2018interpretability}. These could potentially be useful in reconciling a text prompt with a produced image. However, additional explainability techniques produce their own complexities. For example, image generation models might heavily rely on bias between n-grams in an input prompt phrase and a highly abstract concept in a produced image, and such bias might not be easily understandable to humans. Recent papers (e.g.,~\cite{huang2022conceptexplainer}) explore visualization-drive strategies for analyzing these sorts of algorithms, and their techniques could be incorporated into tools like LINGO that expand to additional multimedia-based prompting.
\vspace{-1em}

\section{Conclusion}

We present LINGO, a novel visual analytics interface that supports an effective, task-driven workflow to (1) help identify bias in natural language task instructions, (2) alter (or create) task instructions to reduce bias, and (3) evaluate pre-trained model performance on debiased task instructions. Our evaluations demonstrate how LINGO promotes the creation of more difficult tasks for pre-trained language models that contain higher linguistic diversity and lower instruction bias. We also discuss how visual analytics tools like LINGO can reduce the effort involved in prompt creation, including with constraints like increasing task diversity and lowering instructional bias, across additional domains.

\section{Acknowledgements}

This research was supported in part by the U.S. National Science Foundation through grant DUE-2216452.
\bibliographystyle{eg-alpha-doi}  
\bibliography{00_main}        
\end{document}